\documentclass[
preprint,
 amsmath,amssymb,
 aps,
]{revtex4-2}

\usepackage{graphicx}% Include figure files
\usepackage{dcolumn}% Align table columns on decimal point
\usepackage{bm}% bold math
\usepackage[colorlinks=true,allcolors=blue]{hyperref}
\usepackage{float}
\usepackage{xcolor}
\usepackage{natbib}
\begin{document}

\title{Pattern Formation of Freezing Infiltration in Porous Media}

\author{Nathan D. Jones}
\email{ndjones@caltech.edu}
\author{Adrian Moure}
\email{amoure@caltech.edu}
\author{Xiaojing Fu}
 \email{rubyfu@caltech.edu}
\affiliation{
 Department of Mechanical and Civil Engineering, California Institute of Technology, 1200 E California Blvd, Pasadena CA, 91125
}
\date{\today}

\begin{abstract}
Gravity-driven infiltration of liquid water into unsaturated porous media can be a spatially heterogeneous process due to the gravity fingering instability.
When such infiltration occurs in a subfreezing porous medium, liquid water can readily freeze, leading to both the removal of liquid water available for transport and a reduction in local permeability.
As a result of the coupling between gravity fingering and freezing, macroscopic frozen structures can form that record the shape and history of the wetting front.
These structures have been observed in the field in terrestrial snowpack and glacial firn layers and are believed to have profound impacts on how liquid water and its accompanying thermal content distribute during infiltration.
However, a more detailed physics-based understanding of freezing infiltration has been missing.
In this work, we use a thermodynamic nonequilibrium infiltration model to investigate the emergence of refrozen structures during water infiltration into an initially homogeneous and subfreezing porous medium.
From scaling analysis, we recover the relevant nondimensional groups that govern the physics of the freezing infiltration process.
We identify two key mechanisms caused by freezing that reduce the effective infiltration rate, calculated as the maximum depth of infiltration per elapsed time.
In the first mechanism, the effective infiltrate rate decreases because a portion of the liquid water is consumed due to freezing, and such effect can be well quantified by the freezing Damk\"ohler number.
For the second mechanism, we report on a new phenomenon termed secondary fingering, where new flow paths are established in between the primary infiltration channels.
We find that secondary fingering reduces the degree of flow channelization and thus weakens the effective rate of infiltration via flow field homogenization.
% \begin{description}
% \item[Usage]
% Secondary publications and information retrieval purposes.
% \item[Structure]
% You may use the \texttt{description} environment to structure your abstract;
% use the optional argument of the \verb+\item+ command to give the category of each item. 
% \end{description}
\end{abstract}

%\keywords{Snow, meltwater, infiltration, test keyword}%Use showkeys class option if keyword display desired
\maketitle

%\tableofcontents
\newpage
%%%%%%%%%%%%%%%%%%%%%%%%%%%%%%%%%%%%%%%%%%%%%%%%%%%%%%%%%%%%%%%
%--------------------------SECTION I--------------------------%
%%%%%%%%%%%%%%%%%%%%%%%%%%%%%%%%%%%%%%%%%%%%%%%%%%%%%%%%%%%%%%%
\section{Introduction}\label{sec:secIntro}
When a viscous fluid infiltrates into an initially dry porous medium under gravity, an instability can readily occur that results in the formation of preferential flow pathways.
In the absence of macro pores or other types of heterogeneities in the porous medium \cite{bouma1978case,beven1982macropores}, pore-scale competition amongst gravitational, viscous, and capillary forces at the wetting front gives rise to this instability.
This phenomenon, often referred to as the gravity fingering instability, is widely observed in the field \cite{ritsema1996predicted, campbell_role_2006,clerx_situ_2022} and studied in laboratory experiments \cite{glass_physics_1996,selker1992wetting,yao2001stability,wei_morphology_2014} for various porous media such as sand, soil, or snow.
Preferential flow in unsaturated media will enhance the effective infiltration rate of the wetting-phase fluid compared to stable flow \cite{marsh1984wetting}, and has shown to accelerate the arrival of solutes or contaminants \cite{glass1989preferential,kung2000impact} into deeper regions of the domain.

\begin{figure}
\centering
\includegraphics{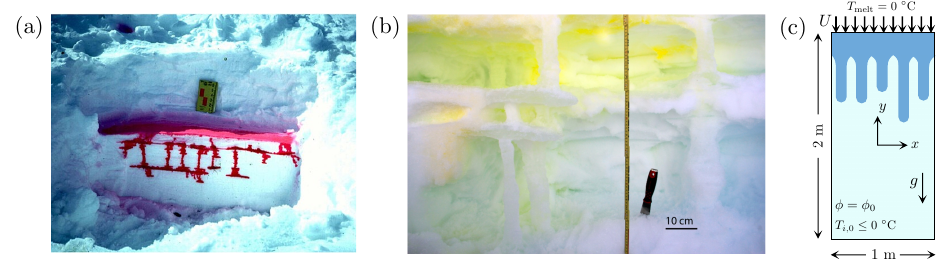}
\caption{\label{fig:fig1} Preferential flow during infiltration of water into subfreezing porous ice: (a) Photograph of a red dye tracer experiment from Ref. \cite{marsh1984wetting} a snow pit exhibiting preferential flow. Photograph courtesy of Philip Marsh. (b) Field image from Ref. \cite{miller_hydrology_2020} of excavated refrozen structures created after preferential infiltration in snow. (c) Problem setup for simulations presented in this study. Liquid water at $T_\mathrm{melt}=0~^\circ$C infiltrates through the top of the domain and interacts with the subfreezing porous ice.}
\end{figure}

Gravity fingering has recently gathered a renewed interest in the field of snow and firn hydrology.
In this context, water and its associated thermal content enter the snowpack from radiation-induced surface melting or rain-on-snow events \cite{wurzer2017modelling,abermann2019large} and actively modify the thermal budget and permeability structure of the porous ice.
Experiments of gravity-driven water percolation both in the field, as shown in Fig. \ref{fig:fig1}(a), and in the laboratory in 3D snow columns have demonstrated that the process is intrinsically unstable, giving rise to preferential pathways \cite{kinar_measurement_2015, waldner_effect_2004, katsushima2013experimental}.
Meanwhile, as snow exists below the freezing point of water, heat transfer and phase change processes readily occur during the infiltration process.
In contrast to isothermal gravity fingering, here the effect of phase change will hinder the vertical infiltration rate of meltwater by converting the amount of liquid water available for transport into solid ice and consequently lowering the local permeability in the snow.
This refreezing of melt within the snowpack forms cylindrical frozen structures known as \textit{ice pipes} on the order of centimeters in width at the site of preferential flow pathways \cite{woo1982basal,williams2000ice,miller_hydrology_2020,bouchard2024impact}.
Additionally, during contact with discrete snow layers below the surface, vertical flow can be stopped and redirected laterally, forming horizontal frozen structures commonly referred to as ice lenses \cite{clerx_situ_2022}.
Evidence of these refrozen structures has been widely observed in the field \cite{benson1960stratigraphic,echelmeyer1992surficial}, both from excavated snow pits [Fig. \ref{fig:fig1}(b)] as well as radar detection \cite{heilig2018seasonal,culberg2021extreme} for larger-scale frozen structures such as ice lenses or slabs.
The formation of refrozen structures is known to affect the thermal budget of the snowpack due to latent heat release \cite{van2012simulating,van2016changing}, and the lasting structures affect the bulk hydraulic, thermodynamic, and structural properties of the snowpack. 

%FIGURE 1

Motivated by these observations, there have been a series of experimental and modeling efforts to better understand the infiltration process of water through permeable ice such as snow or firn.
Experimental studies that investigate gravity-driven preferential flow in porous ice typically operate at isothermal conditions \cite{katsushima2013experimental,avanzi_observations_2016,clerx_situ_2022} in which no phase change is permitted to occur.
Additionally, due to the opacity of porous ice, directly imaging the macroscopic infiltration process is difficult, and requires the dissection of experiments \textit{a posteriori} to investigate the flow patterns \cite{avanzi_observations_2016} or, more recently, the employment of non-intrusive techniques such as neutron radiography \cite{lombardo2023method} or magnetic resonance imaging \cite{adachi2020application,yamaguchi2023novel}.
Many numerical studies have been performed that investigate the formation of refrozen structures (in particular, ice lenses) using one-dimensional models \cite{de2013investigating, wever2016simulating,meyer2017continuum,webb2021two} without considering the flow instability.
Preferential flow in porous ice has also been investigated through numerical simulations, but these studies mainly consider isothermal conditions \cite{hirashima2014multi,hirashima2019wet,leroux2019simulation,leroux2020simulation} or are limited by the assumption of local thermodynamic equilibrium \cite{illangasekare1990modeling,leroux2017modelling}.
Despite the efforts mentioned above, a systematic investigation of the mechanical details of infiltration into subfreezing porous media and the resulting mass transfer dynamics is still lacking.

In this work, we investigate gravity-driven unstable infiltration of liquid water into an initially dry porous medium composed of ice and air.
We use a recently developed model \cite{moure_thermodynamic_2023} that couples a two-phase flow model of unstable infiltration with nonequilibrium thermodynamics of ice-water heat transfer and phase change. 
In this model, water flow is prescribed with a Darcy velocity that takes contributions from gravity, capillary pressure, and macroscopic surface tension effects.
The model captures phase change between liquid water and ice by assigning independent temperature fields to the water and ice phases and thus allowing local thermodynamic non-equilibrium (LTNE).
The LTNE assumption allows us to impose phase change rates based on thermodynamic properties and local temperatures of the porous media instead of fixed rates of phase change.
We perform high-resolution numerical simulations of the model on a 2D rectilinear domain and focus our analysis on the nonlinear regime past the onset of the fingering instability.
Our simulations capture the formation of vertical refrozen structures during infiltration over a wide range of flow rates and thermal conditions.
Quantitative analysis of the simulations shows that the effective infiltration rate coupled with phase change can be well predicted by the freezing Damk\"ohler number.
We also show that increased rates of both freezing and water flux can decrease the heterogeneity of the flow field by forming additional preferential pathways.
We report on a new phenomenon we term as ``secondary fingering'', where new infiltration channels form between pre-established ones.
This phenomenon occurs only under the regime of high rates of both freezing and flow, and arises primarily due to the permeability reduction from freezing.
The secondary fingering process is in contrast to classical gravity-driven infiltration, in which the infiltrating fluid tends to travel through pre-existing pathways.

%%%%%%%%%%%%%%%%%%%%%%%%%%%%%%%%%%%%%%%%%%%%%%%%%%%%%%%%%%%%%%%%
%--------------------------SECTION II--------------------------%
%%%%%%%%%%%%%%%%%%%%%%%%%%%%%%%%%%%%%%%%%%%%%%%%%%%%%%%%%%%%%%%%
\section{Governing Equations}\label{sec:secMethods}
We consider a porous matrix of ice, water, and air, in which the volume-averaged quantities water saturation $S(\mathbf{x},t)$ and porosity $\phi(\mathbf{x},t)$ determine the phase composition throughout the domain. The volumetric liquid water content (LWC) is defined as $\mathrm{LWC}=\phi S$ and the solid fraction of ice is $1-\phi$. 
Flow of water is governed by a Darcy velocity $\boldsymbol{u}(\mathbf{x},t)$. 
The temperature of the water and ice phases are  $T_w(\mathbf{x},t)$ and $T_i(\mathbf{x},t)$, respectively. 
The full details of this model can be found in Ref. \cite{moure_thermodynamic_2023}. 
Here, we provide a brief summary of the governing equations and their underlying assumptions.

\noindent The conservation equations for ice and water mass take the form:
\begin{gather}
    \label{eqn:iceMassConservation}
    \frac{\partial(1-\phi)}{\partial t}=-R_m W_\mathrm{SSA}\left(T_\mathrm{int}-T_\mathrm{melt}\right), \\
    \label{eqn:waterMassConservation} 
    \frac{\partial(\phi S)}{\partial t}+\nabla\cdot\boldsymbol{u}=\frac{\rho_i}{\rho_w} R_m W_\mathrm{SSA}(T_\mathrm{int}-T_\mathrm{melt}), 
\end{gather}
\noindent where $\rho_i$ and $\rho_w$ are the densities of ice and water, respectively, $T_\mathrm{int}$ is the temperature of the ice-water interface, and $T_\mathrm{melt}=0$ $^\circ$C is the melting point of ice. 
The phase change rate coefficient $R_m$ is defined as $R_m=c_{p,w}/(L_\mathrm{sol}\beta_\mathrm{sol})$, where $c_{p,w}$ is the specific heat capacity of liquid water, $L_\mathrm{sol}$ is the latent heat of solidification of ice, and $\beta_\mathrm{sol}$ is the kinetic attachment coefficient for ice growth from liquid water \cite{libbrecht_physical_2017}. 
The wet specific surface area $W_\mathrm{SSA}$ represents the ice-water interfacial area per unit volume:
\begin{equation}
    \label{eqn:Wssa}
    W_\mathrm{SSA}=S\frac{\mathrm{SSA}_0}{\phi_0\mathrm{ln}(\phi_0)}\phi\mathrm{ln}(\phi),
\end{equation}
where $\mathrm{SSA}_0$ is the initial specific surface area. 
The interfacial temperature $T_\mathrm{int}$ is an averaged pore-scale quantity that is defined from the volume integrated Stefan condition:
\begin{equation}
\label{eqn:StefanCondition}
    \frac{T_\mathrm{int}-T_\mathrm{melt}}{L_\mathrm{sol}/c_{p,w}}=-\beta_\mathrm{sol} \overline{v_n},
\end{equation}
where $\overline{v_n}$ is the volume averaged velocity of the fluid-solid interface and is taken to be positive for freezing.
Following \citet{moure_thermodynamic_2023}, the interface temperature takes the form:
\begin{equation}
        T_\mathrm{int}(T_i,T_w)=\frac{\frac{k_i}{r_i}T_i+\frac{k_w}{r_w}T_w}{\frac{c_{p,w}\rho_w}{\beta_\mathrm{sol}}+\frac{k_i}{r_i}+\frac{k_w}{r_w}},
\end{equation}
where $k_i$ and $k_w$ are the thermal conductivity of the ice and water phases, and $r_i$ and $r_w$ are thermal length scales of the ice and water phases, respectively.
The parameters $r_i$ and $r_w$ represent thermal diffusion lengths that are determined from pore-scale simulations of the generalized Stefan problem for solidification (further details in \citet{moure_thermodynamic_2023}).
Note that the right-hand side expressions in Eqs. \eqref{eqn:iceMassConservation} and \eqref{eqn:waterMassConservation} differ by a factor of $\rho_i/\rho_w$ and have opposite signs, which ensures mass conservation during phase change.

The model assumes local thermodynamic nonequilibrium by tracking the temperatures of the ice and water phases independently:
\begin{gather}
\label{eqn:iceThermal}
    \frac{\partial [(1-\phi) T_i]}{\partial t} = \nabla\cdot \left(D_i (1-\phi) \nabla T_i \right) + W_\mathrm{SSA}D_i\frac{T_\mathrm{int}-T_i}{r_i}, \\[0.5em]
    \label{eqn:waterThermal}
    \frac{\partial (\phi S T_w)}{\partial t} + \nabla\cdot \left(\boldsymbol{u} T_w\right) = \nabla\cdot \left(D_w \phi S \nabla T_w \right) + W_\mathrm{SSA}D_w\frac{T_\mathrm{int}-T_w}{r_w},
\end{gather}
where and $D_i$ and $D_w$ are the thermal diffusivities of ice and water, respectively. 

\begin{figure}
\centering
\includegraphics[width=0.85\linewidth]{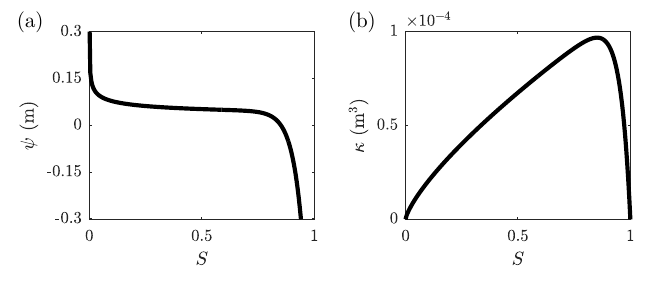}
\caption{\label{fig:fig2} Capillary pressure $\psi$ and the capillary pinning function $\kappa$ used in this work. Here, we plot Eq. \eqref{eqn:psi} and \eqref{eqn:kappa} with $h_\mathrm{cap}=2.5$ cm, $\alpha=4$, $\beta=22$, and $\nu_e=1$. Parameters are calibrated from experimental measurements from Ref. \cite{katsushima2013experimental}.}
\end{figure}

To capture unstable gravity-driven infiltration in porous media, here we adopt the unsaturated flow model proposed by Cueto-Felgueroso and Juanes \cite{cueto2008nonlocal,cueto2009phase,beljadid_continuum_2020}:
\begin{equation}
\label{eqn:velocity}
    \boldsymbol{u} = -K_s(\phi)k_r(S)\nabla\Pi(S),
\end{equation}
where $K_s(\phi)$ is the saturated hydraulic conductivity, $k_r(S)$ is the relative permeability, and $\Pi(S)$ is the total flow potential.
We use an empirical expression for the snow hydraulic conductivity \cite{calonne20123}:
\begin{equation}
    \label{eqn:Ks}
    K_s(\phi)=3\left(\frac{d_i}{2}\right)^2\frac{\rho_w g}{\mu_w}\mathrm{exp}\left[-0.013\rho_i(1-\phi)\right],
\end{equation}
where $d_i$ is the average ice grain diameter, $g$ is the gravitational acceleration, and $\mu_w$ is the dynamic viscosity of water, all of which we assume to be constant. The relative permeability takes the form of a convex function of saturation \cite{brooks1966properties}:
\begin{equation}
    \label{eqn:kr}
    k_r(S)=\left(\frac{S-S_r}{1-S_r}\right)^a,
\end{equation}
where $S_r$ is the irreducible saturation and the exponent $a>1$ is dependent on the porous media. Here, we take $S_r=1\times10^{-3}$ and $a=5$. The expression for the flow potential is \cite{cueto2008nonlocal,cueto2009phase}:
\begin{equation}
\label{eqn:potential}
    \Pi(S)=y-\psi(S) - \sqrt{\kappa}\nabla\cdot(\sqrt{\kappa}\nabla S),
\end{equation}
where $y$ is the vertical coordinate (increasing with height), $\psi(S)$ is capillary suction derived from Leverett scaling, and $\kappa(S)$ is the capillary pinning function:
\begin{gather}
\label{eqn:psi}
    \psi(S)  =  h_\mathrm{cap} S^{-\frac{1}{\alpha}} \left\{ 1-\mathrm{exp} \left[ \beta(S-\nu_e) \right] \left( 1+\beta\frac{\alpha}{\alpha-1} S \right) \right\},\\[0.5em]
    \label{eqn:kappa}
    \kappa(S) = h_\mathrm{cap}^2\int_0^S \psi(S) \,\mathrm{d}S = h_\mathrm{cap}^3 \frac{\alpha}{\alpha-1} S^{\frac{\alpha-1}{\alpha}} \left\{ 1-\mathrm{exp} \left[ \beta(S-\nu_e) \right] \right\},
\end{gather}
where $h_\mathrm{cap}$, $\alpha$, $\beta$, and $\nu_e$ are constants that are calibrated from water retention curves for a given porous material \cite{yamaguchi2010water,yamaguchi2012dependence,katsushima2013experimental}. 
These functions along with the values for these parameters are shown in Fig. \ref{fig:fig2}.
Note that in the case of $\kappa=0$, the classical Richards equation for unsaturated flow in porous media is recovered. 

% FIGURE 2

\section{Problem Setup and Scaling analysis}\label{sec:secMethodsScaling}
\begin{figure}
\centering
\includegraphics{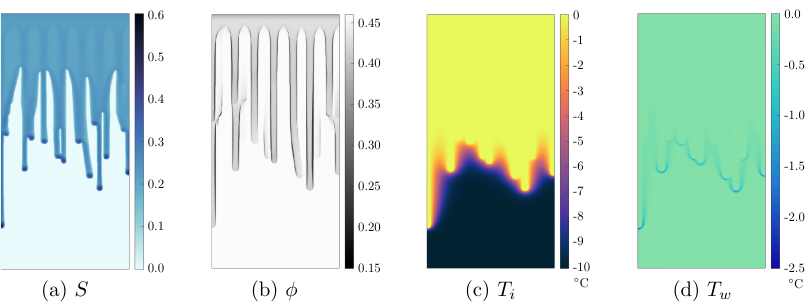}
\caption{\label{fig:fig3} Snapshots of problem unknowns at $t=200$ min after continuous infiltration for: (a) saturation $S$, (b) porosity $\phi$, (c) ice temperature $T_i$, and (d) water temperature $T_w$. For this simulation, $U=27$ mm/h, $T_{i,0}=-10$ ºC, $L_x=1$ m, and $L_y=2$ m.
Porosity is decreased locally during freezing, leaving behind vertical frozen structures that show the history of the gravity fingers. Thermal equilibrium in the ice phase temperature $T_i$ is established behind the fingering front, and phase change does not continue to occur in this region. The water temperature $T_w$ also reaches thermal equilibrium behind the wetting front and exhibits local undercooling at the fingering front where phase change is occurring.}
\end{figure}

We apply the above model to the problem of uniform infiltration in 2D, in which we introduce liquid water into a subfreezing 2D porous medium through the top boundary at a constant flow rate $U$, as illustrated in Fig. \ref{fig:fig3}.
The incoming water temperature is prescribed as $T_\mathrm{melt}$, and the initial temperature of the porous ice is $T_{i,0}$.
A no-flux boundary condition is imposed on the lateral boundaries and bottom boundary of the domain for both flow and thermal energy.
We identify the following characteristic scales in our system: $|\Delta T_0|=|T_{i,0}-T_\mathrm{melt}|$ is the characteristic temperature, $U$ is the characteristic velocity, $L=\mathrm{SSA}_0^{-1}$ is the length scale, and hence the product $t_c=(\mathrm{SSA}_0U)^{-1}$ is the characteristic time scale.
We use $\mathrm{SSA}_0^{-1}$ as the length scale as it provides a more appropriate volume-averaged measurement of the average grain size.
Note that $\mathrm{SSA}_0^{-1}\sim d_i$ for spherical ice grains. 
The nondimensional equations for mass and thermal conservation of each phase then take the following form:
\begin{align}
    \label{eqn:iceMassNondimensional}
    \frac{\partial\phi}{\partial t} &= \mathrm{Da} W_\mathrm{SSA}(T_\mathrm{int}-T_\mathrm{melt}), \\[0.5em]
\label{eqn:waterMassNondimensional}
    \frac{\partial (\phi S)}{\partial t}+\nabla\cdot\boldsymbol{u}&=\frac{\rho_i}{\rho_w} \mathrm{Da} W_\mathrm{SSA}(T_\mathrm{int}-T_\mathrm{melt}), \\[0.5em]
\label{eqn:iceThermalNondimensional}
    \frac{\partial [(1-\phi) T_i]}{\partial t} &= \mathrm{Pe}_i^{-1}\left[\nabla\cdot \left( (1-\phi) \nabla T_i \right) + W_\mathrm{SSA}(T_\mathrm{int}-T_i)\right], \\[0.5em]
\label{eqn:waterThermalNondimensional}
    \frac{\partial (\phi S T_w)}{\partial t} + \nabla\cdot \left(\boldsymbol{u} T_w\right) &= \mathrm{Pe}_w^{-1}\left[\nabla\cdot \left(\phi S \nabla T_w \right) + W_\mathrm{SSA}(T_\mathrm{int}-T_w)\right], 
\end{align}

% FIGURE 3

\noindent The Damk\"{o}hler number Da sets the ratio between the rate of phase change and the rate of advection: 
\begin{equation}
\label{eqn:Damkohlerdefinition}
    \mathrm{Da}=\frac{c_{p,w}|\Delta T_0|}{L_\mathrm{sol}U\beta_\mathrm{sol}}.
\end{equation}
Applying Eq. \eqref{eqn:StefanCondition} to the above definition allows us to rewrite Da as the ratio between the freezing velocity at the ice-water interface, $\overline{v_n}$, and the rate of advection: 
\begin{equation}
\label{eqn:Damkohlerdefinition2}
    \mathrm{Da}=\frac{\mathrm{Ste}}{U\beta_\mathrm{sol}}=\frac{\overline{v_n}}{U}.
\end{equation}
Here, $\mathrm{Ste}$ is the Stefan number, which sets the ratio between the amount of sensible heat to the latent heat of solidification: 
\begin{equation}
    \mathrm{Ste}=\frac{c_{p,w}|\Delta T_0|}{L_\mathrm{sol}}.
\end{equation}
As Da increases, the rate of freezing becomes comparable to the rate of infiltration.
The thermal P\'eclet numbers associated with ice and water set the ratio between advection and thermal diffusion:
\begin{align}
    \mathrm{Pe}_i=UL/D_i,& &\mathrm{Pe}_w=UL/D_w.
\end{align}
The thermal P\'eclet numbers control how quickly heat is transferred by diffusion both ahead of the infiltration front as well as in the spanwise direction.
When Pe is sufficiently small, thermal diffusion dominates and any thermal nonequilibrium is due to the kinetic limitations of phase change.
As Pe$_i$ and Pe$_w$ differ only by the thermal diffusivity of each phase, we refer to $\mathrm{Pe}=\mathrm{Pe}_i$ in this work.

The above scaling analysis introduces the nondimensional groups that control the heat transfer and phase change dynamics of the problem.
The analysis assumes the natural length scale of the porous media ($\mathrm{SSA}_0^{-1}$), which correlates with the grain size.
However, the original isothermal problem of gravity-driven infiltration bears its own scaling analysis that yields additional nondimensional groups that control the length scale and onset of the fingering pattern.
Applying the same scaling analysis to Eqs. \eqref{eqn:velocity} and \eqref{eqn:potential}, we arrive at the nondimensional equations for velocity and flow potential:
\begin{gather}
    \label{eqn:flowNondimensional}
    \boldsymbol{u} = -\frac{K'_s(\phi)}{R_s}k_r\nabla\Pi', \\[0.5em]
    \label{eqn:potNondimensional}
    \Pi'=y'-N_\mathrm{gr}^{-1}\psi'(S)-N_\mathrm{nl}\left(\sqrt{\kappa'}\nabla\cdot(\sqrt{\kappa'}\nabla S)\right),
\end{gather}
where $\psi'(S)=\psi(S)/h_\mathrm{cap}$, $\kappa'=\kappa/h_\mathrm{cap}^3$, and $N_\mathrm{gr}$ and $N_\mathrm{nl}$ are the gravity number and nonlocal number, respectively.
Note that in the absence of phase change, $K_s$ is constant, and $U/K_s$ is defined as the flux ratio $R_s$ \cite{glass_physics_1996,wang1998prediction,cueto2008nonlocal}.
The influence of these two additional nondimensional groups on the stability of the flow pattern is investigated in Ref. \cite{cueto2009stability}, and they are defined:
\begin{align}
    N_\mathrm{gr}=\frac{L}{h_\mathrm{cap}},& &N_\mathrm{nl}=\frac{\Gamma}{\rho g L^3},
\end{align}
where $\Gamma$ is a coefficient in the nonlocal free energy potential that is associated with an effective surface tension of the wetting front interface \cite{weitz1987dynamic,cueto2009phase}.
Following Ref. \cite{beljadid_continuum_2020}, we take $\Gamma=\rho g \kappa$, and recognizing that $\kappa\sim h^3_\mathrm{cap}$  from Eq. \eqref{eqn:kappa}, we arrive at the relation:
\begin{equation}
    N_\mathrm{nl}=N_\mathrm{gr}^{-3}.
\end{equation}
Increasing $N_\mathrm{gr}$ has been shown to decrease the intrinsic length scale of the gravitational fingering instability, leading to thinner fingers.
In addition to the hydraulic properties of the porous media such as the average grain size diameter $d_i$ or the capillary height $h_{cap}$, both the flux ratio $R_s$ and the initial saturation $S_0$ of the porous media have been shown to influence the onset and nonlinear regimes of the pattern formation process \cite{glass1989wetting,cueto2008nonlocal,cueto2009stability}.

As we focus on the effects of the supplied flux $U$ and the rate of freezing on the pattern formation process, in this work we do not vary $N_\mathrm{gr}$ nor $N_\mathrm{nl}$.
Although the entry flux $U$ has been shown to influence the size and spacing of fingers in an isothermal scenario, in this work, we explore a more narrow range of flux ratios between $1.2\times10^{-4}$ to $5.3\times10^{-4}$ ($U$ between 6.3 to 27 mm/h), and thus the variability of the finger width and spacing is negligible.
Larger input fluxes may result in significantly wider fingers in these simulations, however here we chose the range of $U$ to reflect appropriate values of surface melt rates for natural snowpack reported in the literature \cite{hurkamp2017influence}.

%%%%%%%%%%%%%%%%%%%%%%%%%%%%%%%%%%%%%%%%%%%%%%%%%%%%%%%%%%%%%%%%%
%--------------------------SECTION III--------------------------%
%%%%%%%%%%%%%%%%%%%%%%%%%%%%%%%%%%%%%%%%%%%%%%%%%%%%%%%%%%%%%%%%%
\section{Simulation Setup and Numerical Methods}\label{sec:secNumerical}
We conduct high-resolution numerical simulations of the model to investigate the dynamics and patterns for flow and porosity structures during water infiltration into subfreezing porous ice.
The simulations are performed on a 2D rectilinear domain of 1 m $\times$ 2 m ($L_x\times L_y$) as shown in Fig. \ref{fig:fig1}(c) and Fig. \ref{fig:fig3}.
Initially, the domain is composed of dry porous ice with a uniform initial ice temperature $T_{i,0}\leq 0$ $^\circ$C, uniform porosity $\phi(\mathbf{x},0)= 0.456$, and average ice grain diameter $d_i=1.5$ mm which mimics the hydraulic properties of coarse snow.
We assume the porous media is initially at the irreducible saturation $S(\mathbf{x},0)= S_r=1\times 10^{-3}$. 
The initial specific surface area is estimated from Ref. \cite{domine_parameterization_2007} to be $\mathrm{SSA}_0=3514$ mm$^{-1}$.
We approximate the kinetic attachment coefficient for ice growth from Ref. \cite{libbrecht_physical_2017} as $\beta_\mathrm{sol}=800$ s/m.

To explore the effect of the relative strength of water freezing rate on the overall infiltration pattern, we systematically vary the surface meltwater flux $U$ and the initial ice phase temperature $T_{i,0}$. 
We vary $U$ between 6.3--27 mm/h to explore flow rates near values reported in Ref. \cite{hurkamp2017influence} for natural systems and consider initial ice temperatures between 0 to -20 $^\circ$C.
The domain has $400\times800$ ($N_x\times N_y$) mesh elements, with the exception of simulations presented in Sec. \ref{sec:secResultsVert}, in which we use a finer mesh resolution.
Equations \eqref{eqn:iceMassConservation}, \eqref{eqn:waterMassConservation}, \eqref{eqn:iceThermal}, \eqref{eqn:waterThermal}, and \eqref{eqn:velocity} are solved using isogeometric analysis \cite{hughes2005isogeometric}---a finite element method that implements B-splines as basis functions.
The time advancement scheme is a generalized--$\alpha$ method with an adaptive time step based on the number of Newton-Raphson iterations.
To implement the above methods, we develop a code using the open source libraries PETSc \cite{petsc-web-page} and PetIGA \cite{PetIGA}.

%%%%%%%%%%%%%%%%%%%%%%%%%%%%%%%%%%%%%%%%%%%%%%%%%%%%%%%%%%%%%%%%
%--------------------------SECTION IV--------------------------%
%%%%%%%%%%%%%%%%%%%%%%%%%%%%%%%%%%%%%%%%%%%%%%%%%%%%%%%%%%%%%%%%
\section{Results}\label{sec:secResults}
%----------------SECTION IV A----------------
\subsection{Impact of freezing on infiltration depth}\label{sec:secResultsCollapse}
\begin{figure}
\centering
\includegraphics[width=\linewidth]{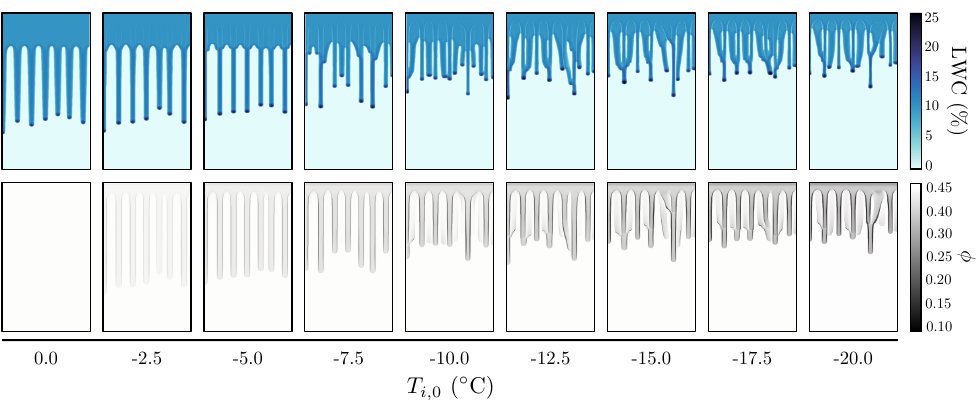}
\caption{\label{fig:fig4} Snapshots of liquid water content (top) and porosity (bottom) at $t_\mathrm{f}=212$ min for varying $T_{i,0}$.  All simulations have the same imposed flux $U=27$ mm/h.}
\end{figure}

Freezing can directly hinder gravity-driven infiltration through two mechanisms.
In the first mechanism, mass in the liquid phase is lost as it is converted into ice at the site of freezing and thus the amount of total water available for transport is reduced.
In the second mechanism, an increase in the ice volume fraction leads to a reduction in local permeability which lowers the Darcy velocity of infiltration.
Both of these effects will slow the effective rate of infiltration of the wetting phase into the porous medium compared to the isothermal problem without freezing.
In Fig. \ref{fig:fig4}, we demonstrate this effect by plotting the liquid water content LWC and porosity $\phi$ at a fixed time ($t_\mathrm{f}=212$ min) after constant infiltration of $U=27$ mm/h for varying $T_{i,0}$.
With decreasing $T_{i,0}$, the wetting front extends shallower into the domain; at the same time, the local porosity reduction due to freezing becomes more pronounced.

% FIGURE 4

\begin{figure}
\centering
\includegraphics{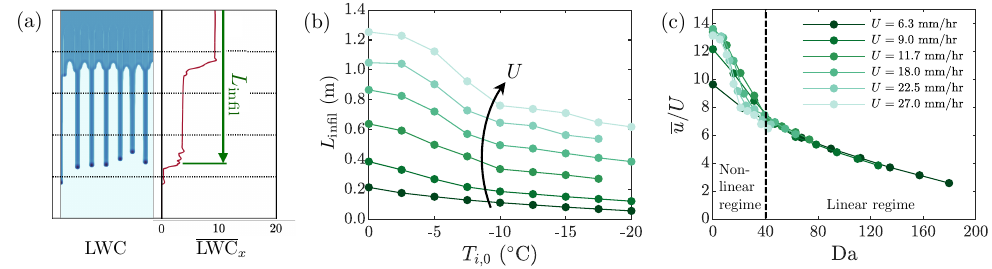}
\caption{\label{fig:fig5} Infiltration depths measured at $t_\mathrm{f}=212$ min over different flow and thermal conditions: (a) At a given time, we calculate the liquid water content $\mathrm{LWC}=\phi S$  (left) and compute its horizontal average (right). The effective infiltration depth $L_\mathrm{infil}$ is then computed as the depth where the integrated volume of $\overline{\mathrm{LWC}}_x$ reaches below 2.5 $\frac{\mathrm{m}^3}{\mathrm{m}^2}$. (b) The effective infiltration depth $L_\mathrm{infil}$ at $t_\mathrm{f}=212$ min is controlled by both $T_{i,0}$ and the imposed flux $U$. (c) We define $\overline{u}=L_\mathrm{infil}/t_\mathrm{f}$ and compute the normalized effective infiltration rate $\overline{u}/U$ as a function of $\mathrm{Da}$.}
\end{figure}

To quantify the decrease of the infiltration depth due to freezing, we define the depth of infiltration $L_\mathrm{infil}$ over a fixed time $t_\mathrm{f}$ by measuring the vertical distance from the top of the domain to the front of infiltration using the horizontally-averaged liquid water content $\overline{\mathrm{LWC}}_x=(\overline{\phi S})_x$ as shown in Fig. \ref{fig:fig5}(a).
In Fig. \ref{fig:fig5}(b), we plot $L_\mathrm{infil}$ at $t_\mathrm{f}=212$ min as a function of both $T_{i,0}$ and $U$.
We find that, for a fixed $U$, the depth of infiltration decreases monotonically with decreasing $T_{i,0}$; meanwhile, at a fixed $T_{i,0}$, $L_\mathrm{infil}$ monotonically increases with increasing $U$.
Applying the appropriate scaling to our data and defining the effective infiltration velocity as $\overline{u}=L_\mathrm{infil}/t_\mathrm{f}$, we find that the curves in Fig. \ref{fig:fig5}(b) can be collapsed into a master curve prescribing the normalized effective infiltration velocity $\overline{u}/U$ as a function of Da, as shown in Fig. \ref{fig:fig5}(c).
In particular, the master curve reveals two distinct regimes of infiltration behavior identified by the threshold $\mathrm{Da}=40$.
When $\mathrm{Da}<40$, the rate of freezing is comparable to the flow rate, and the nonlinear interplay between infiltration and freezing leads to complex behavior in $\overline{u}/U$.
In this regime, increasing $\mathrm{Da}$ reduces $\overline{u}/U$ drastically, and the exact relationship between $\overline{u}/U$ and $\mathrm{Da}$ is highly nonlinear and weakly dependent on $U$. 
When $\mathrm{Da}>40$, the rate of freezing dominates the rate of advection, and $\overline{u}/U$ becomes controlled solely by the rate of freezing and can be prescribed by a linear function of $\mathrm{Da}$.

% FIGURE 5

\subsection{Porosity structure of a freezing finger}\label{sec:secResultsVert}
Our simulation results thus far demonstrate that the initial fingering pattern creates heterogeneous regions of low porosity in an initially homogeneous porous medium, as demonstrated in Fig. \ref{fig:fig3}(b) and Fig. \ref{fig:fig4}. 
The reduction in porosity is directly proportional to the rate of freezing and is focused predominantly at the macroscopic ice-water interface where thermal gradients that drive freezing are the strongest (Eq. \eqref{eqn:iceMassConservation}).
In this section, we study the detailed porosity structure of a freezing finger by considering the formation and infiltration of a single finger into unsaturated permeable ice with subfreezing temperature $T_{i,0}$.
We conduct finer resolution simulations than presented in Sec. \ref{sec:secResultsCollapse} and \ref{sec:secResultsSecondary}, using a domain size of 0.1 m $\times$ 0.5 m ($L_x\times L_y$) with 256 $\times$ 1280 elements ($N_x\times N_y$).
For computational efficiency, we impose a thin ``pinning'' layer at the top of the domain with a lower permeability to promote the fingering instability more quickly.
For the hydraulic properties of the porous medium, we use approximated parameter values from Ref. \cite{yamaguchi2010water} for coarse snow.

\begin{figure}[H]
\includegraphics[width=\linewidth]{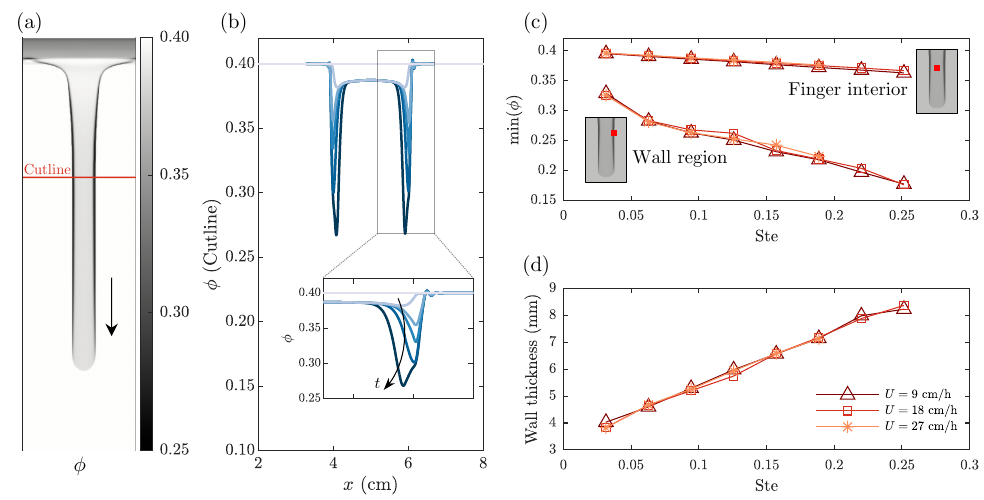}
\centering
\caption{\label{fig:fig6} Freezing infiltration of a single finger: (a) Porosity snapshot of a single finger after infiltrating into porous ice initially at $T_{i,0}=-10$ $^\circ$C. (b) Porosity along a cutline at 0.125 m below the top of the domain at multiple times. Inset shows a zoomed-in region of the frozen wall formed over time until it reaches an equilibrium value. Freezing begins at the macroscopic ice-water interface and propagates inwards where there is available water to freeze. The Stefan number Ste alone controls the (c) minimum porosity for both the frozen wall region (bottom curve) and the interior of the finger (top curve), as well as (d) the thickness of the resulting frozen wall.}
\end{figure}

We perform a suite of these simulations by systematically varying $U$ and $\mathrm{Ste}$ using the same values in Sec. \ref{sec:secResultsCollapse}.
To analyze the results, we consider a cross-section of $\phi$ at $y=0.375$ m [Fig. \ref{fig:fig6}(a)] at a sufficiently late time in which local thermal equilibrium is achieved.
The porosity change inside the finger is small due to the fast downward advection of thermal content which quickly brings the inner region of the finger into thermal equilibrium.
Meanwhile, the outer edges of the finger see sustained and significant freezing, as its nonequilibrium state is maintained by the diffusive heat exchange with the non-infiltrated region [Fig. \ref{fig:fig6}(b)].
We note however, that the ice-water interface eventually approaches thermal equilibrium [Fig. \ref{fig:fig3}(c)] as well, allowing the wall thickness to approach a plateau value as the freezing process halts.

% FIGURE 6

Because we only explore a range of relatively low $U$ that does not influence the finger width, we observe that $U$ only affects the downward finger velocity and does not modify the finger geometry or spanwise heat transfer. 
As a result, the wall thickness and changes in porosity are only affected by Ste. 
The equilibrium configuration of the porosity field exhibits an annulus of a low-porosity region surrounding a still-porous region in which water is still able to percolate through [Fig. \ref{fig:fig6}(a)].
We find that both the wall thickness [Fig. \ref{fig:fig6}(d)] and the porosity of the ice walls [Fig. \ref{fig:fig6}(c)] are only functions of Ste and are independent of the imposed meltwater velocity $U$.
The porosity inside the finger decreases only slightly with Ste, as shown in Fig. \ref{fig:fig6}(c).

\subsection{Formation of secondary fingers}\label{sec:secResultsSecondary}
The reduction in the effective infiltration velocity $\overline{u}$ caused by freezing can be explained, to the first order, by the consumption of melt and the reduction in local permeability.
Here, we highlight an additional mechanism, which we term \textit{secondary fingering}, that weakens the localized nature of the infiltration flow paths and decreases $\overline{u}$.

\begin{figure}
\centering
\includegraphics{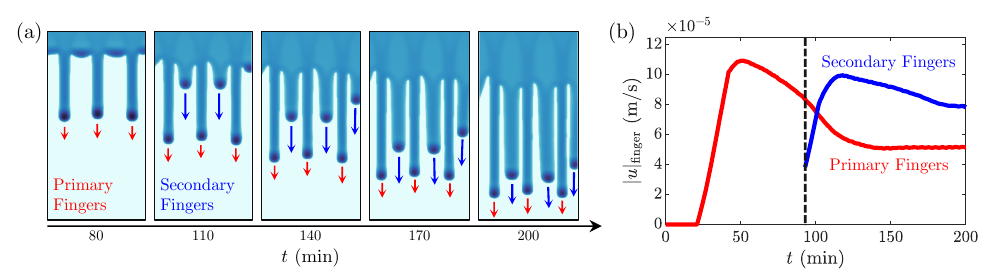}
\caption{\label{fig:fig7} Dynamics of secondary finger formation: (a) Secondary fingers form between primary fingers in regions of local thermal equilibrium. (b) Secondary fingers propagate more quickly than primary fingers, and reduce the infiltration velocity of primary fingers as they redirect water content available for transport. In this simulation, secondary fingers are formed at $t=93$ min (dashed black line). A moving average filter is applied to smooth the average finger velocity curves.}
\end{figure}

During unstable infiltration, an initial set of fingers with roughly uniform spacing forms, which we denote as the \textit{primary} fingers as shown in Fig. \ref{fig:fig7}(a) at $t=80$ min.
The majority of freezing occurs at the ice-water interface of the primary fingers due to high thermal gradients across the interface (Fig. \ref{fig:fig6}).
The latent heat release and accompanying diffusive heat transfer from this freezing warms the unsaturated region between each finger, leading to a region of uniform thermal equilibrium behind the fingering front [Fig. \ref{fig:fig3}(c)].
As the regions between each finger have remained unsaturated, local permeability is higher than the wetted regions that have undergone freezing.
With flow still supplied at the top of the domain, new preferential pathways, which we term the \textit{secondary} fingers, emerge in these regions of higher permeability and local thermodynamic equilibrium, as shown in Fig. \ref{fig:fig7}(a) for $t>80$ min.
Since the conditions for flow are more favorable in these regions, the secondary fingers quickly propagate towards the front near the primary fingers until they reach the front and begin to undergo freezing, as shown in Fig. \ref{fig:fig7}(b).
The formation and propagation of secondary fingers will slow down the overall propagation of the wetting front, as water available for transport is now diverted into the unsaturated regions in between the primary fingers.

% FIGURE 7

\begin{figure}
\centering
\includegraphics{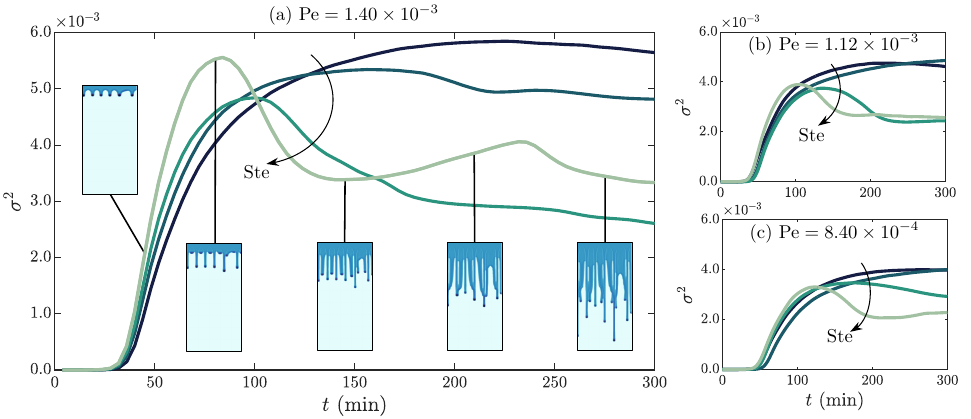}
\caption{\label{fig:fig9} Evolution of the variance $\sigma^2$ of the vertically-averaged saturation $\overline{S}_y$ over time for Ste $\in \{0.031, 0.063, 0.094, 0.126\}$ (increasing in direction of arrows) and for three different Pe. Snapshots in (a) show the saturation field at various times and demonstrate the decrease in $\sigma^2$ due to the formation of secondary fingers, and increases in $\sigma^2$ past $t=150$ min due to secondary finger coalescence.}
\end{figure}

The reduction in effective infiltration rate due to the formation of secondary fingers is also corroborated by a reduction in the degree of channelization of the flow structure.
To demonstrate this, we quantify the lateral flow path distribution by measuring the variance in saturation. 
We introduce $\overline{S}_y(x)$ as the $y$-averaged saturation along the $x$-axis.
We then define the degree of flow channelization as the horizontal variance of $\overline{S}_y$: 
\begin{equation}
    \sigma^2=\langle {\overline{S}}^2_y\rangle-\langle \overline{S}_y\rangle^2.
\end{equation}
For a uniform flow, regardless of saturation overshoot in the $y$-direction, $\overline{S}_y$ will be constant along $x$ and thus $\sigma^2$ will be zero.
In Fig. \ref{fig:fig9}, we demonstrate that, as the formation of secondary fingers redistributes flow into regions where the primary fingers do not visit, the area of the unsaturated region decreases as more channels occupy the domain, and thus $\sigma^2$ decreases.
We also note that before the onset of secondary fingers, $\sigma^2$ increases with increasing $\mathrm{Ste}$ and can even surpass $\sigma^2$ of flow configurations without secondary fingers.
In these scenarios, more freezing occurs at the ice-water interface of each finger for higher Ste (Sec. \ref{sec:secResultsVert}), and thus the effective finger width decreases, leading to sharper gradients in $\overline{S}_y(x)$ and therefore a higher variance.
Additionally, for $\mathrm{Ste}>0.1$ ($T_{i,0}<-7.5$ $^\circ$C), some fingers have the potential to meander and begin to propagate diagonally.
Only the secondary fingers exhibit this meandering behavior, as the porosity structure at the site of their formation is heterogeneous due to freezing from the primary set of fingers.
This can cause slightly more freezing at one side of the meandering finger due to differing local thermodynamic conditions on either side. 
For an initially homogeneous medium, we postulate that this small amount of phase change causes a ``runaway" effect, causing the meandering finger to continue propagating diagonally until it combines with another finger.
The coalescence of two fingers here will also raise $\sigma^2$, as shown in Fig. \ref{fig:fig9}(a) insets.

% FIGURE 9

\begin{figure}[H]
\centering
\includegraphics{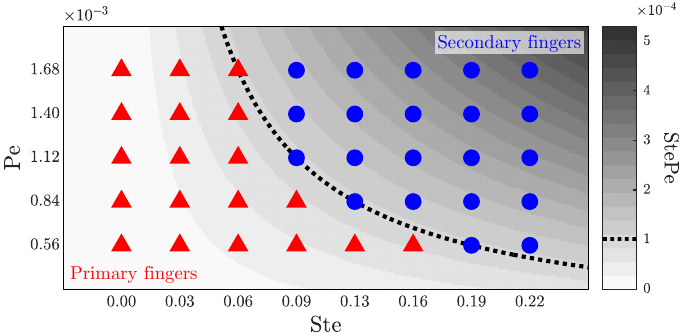}
\caption{\label{fig:fig10} Phase diagram predicting the emergence of secondary fingers. The compounding effects of sufficiently high thermal advection and rates of freezing, as quantified by the nondimensional product StePe, controls the emergence of secondary fingers.}
\end{figure}

In Sec. \ref{sec:secResultsCollapse}, we demonstrate that the competition between the rate of phase change and the rate of advection, or the Damk\"ohler number, controls the normalized effective infiltration rate $\overline{u}/U$. 
Here, we find that the compounding effects of phase change, as controlled by Ste, and thermal advection, as controlled by Pe, predict the formation of secondary fingers.
In Fig. \ref{fig:fig10}, we present a Ste--Pe phase diagram that predicts the emergence of secondary fingers.
Both sufficiently high rates of freezing around the primary fingers (high Ste) as well as sufficient flow available for transport (high Pe) are required to form secondary fingers.
Thus, the product of these nondimensional numbers, $\mathrm{Ste}\mathrm{Pe}$ controls the emergence of secondary fingers.
As shown in Fig. \ref{fig:fig10}, we find that secondary fingers only appear for $\mathrm{Ste}\mathrm{Pe}>10^{-4}$ (dashed curve).

% FIGURE 10

%%%%%%%%%%%%%%%%%%%%%%%%%%%%%%%%%%%%%%%%%%%%%%%%%%%%%%%%%%%%%%%
%--------------------------SECTION V--------------------------%
%%%%%%%%%%%%%%%%%%%%%%%%%%%%%%%%%%%%%%%%%%%%%%%%%%%%%%%%%%%%%%%
\section{Conclusions}\label{sec:secConclusions}
In this paper, we study the effects of freezing on the unstable infiltration of water into porous media.
We use a continuum model \cite{moure_thermodynamic_2023} to describe gravity-driven infiltration coupled with thermal transport and ice-water phase change in an initially unsaturated and subfreezing porous media.
We present high-resolution numerical simulations of the model applied to various thermal conditions and flow rates and identify key nondimensional parameters that govern the patterns of flow and the resulting frozen structures.
Our results demonstrate that the effective infiltration rate can be well characterized by the freezing Damk\"ohler number (Sec. \ref{sec:secResultsCollapse}).
In particular, we identify a regime for $\mathrm{Da}<40$ in which thermal advection and the freezing kinetics are in strong competition, leading to a nonlinear decrease in the effective infiltration rate with increasing $\mathrm{Da}$.
We also identify a linear regime ($\mathrm{Da}>40$) where the effects of freezing dominate the infiltration process and the effective infiltration rate decreases linearly with increasing Da.
In Sec. \ref{sec:secResultsVert}, we focus on the characteristics of vertical structures, or ice pipes, and their formation.
We find that the ice pipes are initially created by freezing at the macroscopic ice-water interface while the interior region of the finger remains porous and continues to facilitate downward transport of water.
Upon future cycles of environmental forcing, this saturated region has the possibility to fully freeze which could result in a solid column of ice, as observed in the field \cite{miller_hydrology_2020,bouchard2024impact}.
In addition to the consumption of liquid water and the decrease in local permeability, we identify a third mechanism---the emergence of secondary fingers---that reduces the effective infiltration rate by diverting water into new flow channels behind the initial fingering front (Sec \ref{sec:secResultsSecondary}).
We find that both sufficient rates of freezing and adequate thermal transport, as described by the nondimensional product $\mathrm{Ste}\mathrm{Pe}$, is required to produce secondary fingers.
By measuring the variance of $\overline{S}_y(x)$, we show that the creation and propagation of secondary fingers reduces the degree of channelization of the flow field.

An important assumption we make in this study is that local freezing does not alter the capillary pressure function $\psi(S)$.
This function, also called the \textit{water retention curve}, has been documented only for a handful of various snow types, but only under isothermal flow conditions \cite{yamaguchi2010water,yamaguchi2012dependence,katsushima2013experimental}.
In freezing flows through porous media, both the volume-averaged ice grain diameter $d_i$ and the ice fraction $(1-\phi)$ would increase at the site of phase change, and the capillary suction force may strengthen or weaken locally and modify the infiltration dynamics.
Future studies involving numerical simulations at the pore-scale and/or laboratory experiments would be valuable in determining how the freezing process affects $\psi(S)$.
Further, as natural snowpack is often a stratified, heterogeneous porous medium, the interaction between gravity-driven flow and discretely layered porous media with varying hydraulic and capillary properties, and in particular the formation of ice lenses, will be a topic of future studies.

\newpage
% \begin{acknowledgments}
% We wish to acknowledge the support of the author community in using
% REV\TeX{}, offering suggestions and encouragement, testing new versions,
% \dots.
% \end{acknowledgments}

\appendix

\newpage

\bibliography{patternForm}% Produces the bibliography via BibTeX.

\end{document}